\documentclass{llncs}
\usepackage{algpseudocode}
\usepackage{algorithm}

\usepackage{xyling}
\usepackage{tikz}  
 \usetikzlibrary{fit}
\usetikzlibrary{trees}
\usetikzlibrary{arrows}
\usepackage{etex}

\usepackage{amsmath}
\usepackage[lofdepth,lotdepth]{subfig}
\usepackage[normalem]{ulem}
\usepackage{colortbl}
\usepackage{listings}
\usepackage{chngcntr}
\usepackage{url}
\usepackage{paralist}
\usepackage{fancybox}
\usepackage{float}
\usepackage{xspace}
\def\fedra{\textsc{Fedra}\xspace}

\newcolumntype{C}[1]{>{\centering\let\newline\\\arraybackslash\hspace{0pt}}m{#1}}
\usepackage[utf8]{inputenc}

\title{\fedra: Query Processing for SPARQL Federations with Divergence}
\author{Gabriela Montoya\inst{1}\inst{2} \and   Hala Skaf-Molli \inst{1} \and Pascal Molli \inst{1} \\ 	Maria-Esther Vidal \inst{3}}

\institute{
LINA-- Nantes University, France \\
\email{\{gabriela.montoya,hala.skaf,pascal.molli\}@univ-nantes.fr}
\and CNRS Unit UMR6241, France\\
\and  Universidad Sim\'on Bol\'{\i}var, Venezuela \\  
\email{mvidal@ldc.usb.ve}
}

\begin{document}
\counterwithout{lstlisting}{section}

\maketitle
\pagestyle{plain}
\vspace{-0.5cm}
 \begin{abstract}

   Data replication and deployment of local SPARQL endpoints improve
   scalability and availability of public SPARQL endpoints, making the
   consumption of Linked Data a reality. This solution requires
   synchronization and specific query processing strategies to take
   advantage of replication. However, existing replication aware
   techniques in federations of SPARQL endpoints do not consider data
   dynamicity. We propose \fedra, an approach for querying federations
   of endpoints that benefits from replication. Participants in \fedra
   federations can copy fragments of data from several datasets, and
   describe them using provenance and views. These descriptions enable
   \fedra to reduce the number of selected endpoints while satisfying
   user divergence requirements. Experiments on real-world datasets
   suggest savings of up to three orders of magnitude.

\end{abstract}

\section{Introduction}

Linked Open Data makes millions of triples available for processing
SPARQL queries through federation of SPARQL endpoints. However,
infrastructure of SPARQL endpoints have intrinsics limitations in
terms of scalability and availability. According to
~\cite{DBLP:conf/semweb/ArandaHUV13}, on 427 public endpoints of
federation only one third have an availability rate above 99\% .

Traditional approaches for improving data availability involves
fragmentation~\cite{verborgh2014web} and
replications~\cite{DBLP:conf/semweb/SaleemNPDH13}. According to
incoming queries, data publishers compute fragments, replicate and
smartly place them on their managed clusters to improve data
availability. Unfortunately, for these approaches, scalability and
availability only relies on resources of data publishers. We advocate
for a vision where data scalability and availability costs should be
supported by the whole federation of linked data relying on existing
standards.

Fragmentation and replication can occur opportunistically among
federation members. One participant can materialize a fragment of
another one for its own purpose and make it available for the whole
federation. A federated query engine that take advantage of such
opportunistic fragmentation and replication can improve general
scalability and availability of linked open data.

\begin{minipage}{0.55\linewidth}
 \begin{lstlisting}[basicstyle=\small\sffamily,language=sparql,caption={DBpedia Query Q},numbers=none,frame=none,columns=fixed,label=queryQ,extendedchars=true,breaklines=true,showstringspaces=false]
SELECT DISTINCT *
WHERE {
  ?city <http://dbpedia.org/ontology/country> ?c .
  ?city <http://dbpedia.org/ontology/department> ?d .
  ?city <http://dbpedia.org/ontology/postalCode> ?pc 
}
\end{lstlisting}
\end{minipage}

Unfortunately, existing federated query engines do not exploit such
opportunities.  To illustrate, consider a DBpedia dataset $d_1$, and a
federation that only accesses a public SPARQL endpoint of
DBpedia. Using FedX~\cite{DBLP:conf/semweb/SchwarteHHSS11}, the
execution of the three-triple pattern query in Listing\ref{queryQ} is
quite simple because the query can be exclusively executed in one
endpoint.  If the same query were executed in a federation with a
mirror of DBpedia ($d_2$), each triple pattern should be executed
against both endpoints, thus query performance
deteriorates. Nevertheless, using the knowledge that $d_2$ is just a
mirror of $d_1$ leads to a simple query strategy that surely will save
execution time. On the other hand, using mirrors raise the issue of
consistency of different fragment and dynamicity of data. Performing
queries in presence of divergent data leads to stale
answers~\cite{DBLP:journals/cacm/GolabRAKL14}.

Recently, Saleem et al. ~\cite{DBLP:conf/semweb/SaleemNPDH13}  propose DAW, 
a framework able to detect
data duplication between datasets and reduce the number of selected
sources.  DAW uses data present in the federation at a given 
time to build indexes for each dataset and is able to detect
quickly data duplication between two datasets.  
In the previous example, FedX with DAW
would contact only one DBpedia endpoint.  However, DAW does not
consider fragments nor source dynamicity, i.e., DAW summaries has to be
recomputed after each change.

In this paper we describe \fedra, a source selection strategy that can
be used in conjunction with existing federated query engine in order
to improve efficiency of queries and data availability. \fedra
considers a set of endpoints $S$ exposing fragments. A fragment is
defined as a SPARQL construct query, the original data source, a value
expressing freshness of the fragment according to original
source. Given a query Q, and divergence threshold, \fedra computes the
minimum set of endpoints to resolve the query. Divergence threshold
allow to control how much stale values can be retrieved in query
results. 

On experimental setup based on DBpedia, we observe that \fedra takes
advantage of opportunistic replication to significantly reduce
execution time of federated queries while bounding stale values even
with simple divergence metrics.

The main contributions of this paper are:
\begin{inparaenum}[\itshape i\upshape)]
\item Endpoint descriptions in terms of SPARQL views, containment relationships between views, data provenance, and timestamps;  
\item \fedra source selection algorithm that reduces the number of selected sources while selected sources divergence is controlled; and 
\item an experimental study that reveals the benefits of both exploiting knowledge encoded in endpoint descriptions and controlling divergence to avoid obsolete results.
\end{inparaenum}

The paper is organized as follows: 
Section~\ref{sec:approach} presents  \fedra and the source selection algorithm.
 Section~\ref{sec:experiments} reports our
experimental study.  Section~\ref{sec:relatedWork} summarizes related
work. Finally, conclusions and future work are outlined in
Section~\ref{sec:conclusion}.
  
\section{\fedra Source Selection for SPARQL Federations with Replication}
\label{sec:approach}

The lack of reliability of SPARQL endpoints is currently forcing
intensive linked data consumers, or linked data application
developers to rely on dataset dumps and local re-installation to
fulfill their own needs. Some data producers such as DBpedia or
MusicBrainz are also providing live update feeds to keep local mirrors
up-to-date and leverage the load on their own SPARQ endpoints.

\begin{figure}
  \centering
  \includegraphics[width=0.95\textwidth]{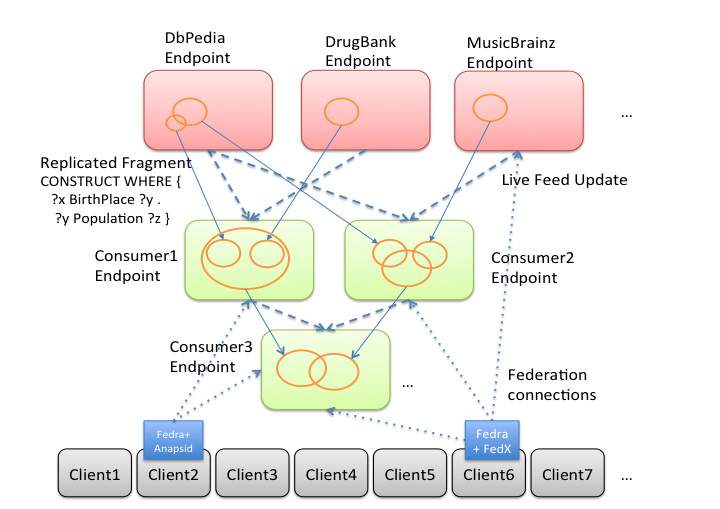}
  \caption{Fedra general approach}
  \label{fig:fedra}
\end{figure}

In \fedra, we generalize this approach by defining and replicating
fragments in an opportunistic way. For example, in
figure~\ref{fig:fedra}, one opportunistic federation may be composed
by public endpoints for datasets
Musicbrainz\footnote{http://dbtune.org/musicbrainz/sparql},
DBpedia\footnote{http://dbpedia.org/sparql}, and
Drugbank\footnote{http://drugbank.bio2rdf.org/SPARQL}, and 3
intensive data consumers. Such deployment allows data producers to
share the load on their own endpoints and data consumers to access
to the knowledge of federated queries issued by the community
concerning these fragments.

Fragments are defined by regular CONSTRUCT SPARQL queries that can be disjoint,
overlapping, or contained with other fragments. Live Update Feeds as in
DBpedia Live, SPARQL Push publish-subscribe~\cite{passant2010sparqlpush}, or just re-executing
fragments queries allow to maintain fragments up-to-date but at
unpredictable rates.

Finally, clients declare original and consumers endpoints in their
federated queries. \fedra compute the minimal set of endpoints where a
given query has to be executed taking into account the freshness on
fragments and delegate the execution of the query to an existing
federated query engine such as FedX or Anapsid~\cite{DBLP:conf/semweb/AcostaVLCR11}.

\fedra will take advantage of locality properties introduced by
opportunistic presence of fragments on the same endpoints to solve the
source selection problem and consequently improve the query execution
time, and data availability. Divergence metrics allow to control
synchronization issues between fragments. Selected aged fragments can
introduce stale values in the query result. In \fedra, clients can express
the divergence tolerance between fragments to impact number of expected stale
values in query results.

Before presenting the problem statement  for the source selection problem \fedra solves, we introduce some definitions:

%\begin{definition}[RDF dataset~\cite{rdfsparqlquery}]
%An RDF dataset represents a collection of graph. Graphs are constituted by triples with subject, predicate and object.
%\end{definition}

%Hereafter, we will simply use the term 'dataset', and the presented approach would overlook the existence of multiple graphs.
%Nevertheless this approach could be easily extended to handle named graphs.

%\begin{definition}[SPARQL endpoint]
%A SPARQL endpoint is a web service that allows to process queries over datasets. An endpoint is defined by
%its Uniform Resource Identifier (URI), protocols and formats supported and the datasets it provides access for.
%\end{definition}

%\begin{definition}[Public endpoint, Data provider]
%Data providers are the entities in charge of integrating data into the linked data cloud.
%To provide access to these data, several data providers opt for deployment and maintenance of 
%a SPARQL endpoint that can be accessed by anyone. Hereafter, these endpoints are 
%called public endpoints.
%\end{definition}

\begin{definition}[Data fragment]
A data fragment is a subset of the triples that constitute a dataset. It can be described using a query
that gives the pattern satisfied by the fragment members, or using the triples contained in it 
at a given time.
\end{definition}

The first description is given in terms of the schema the fragment triples are compliant to, and
the second one is instance dependent. A recent study~\cite{DBLP:conf/esws/KaferAUOH13} has 
shown that the schema of datasets is generally static and have a lot less changes instances, hence
using queries to describe fragments provide a more reliable description. 
%As descriptions, we will use \emph{CONSTRUCT} SPARQL queries that allows to retrieve a 
%dataset fragment, \emph{i.e.}, a subset of the dataset.

\begin{definition}[Replica]
Replicas are fragments that do not represent new data, but that was taken from an existing dataset. 
\end{definition}

\begin{definition}[Replica description]
A replica may be uniquely described by the source from where the replica was made, the query that
when performed in that source returns the replica fragment, the date when the replica was made, and
the SPARQL endpoint that provides access to it.
\end{definition}

%\begin{definition}[Data replicator]
%Data replicators are data consumer that have taken replicas from existing data providers and have
%deployed endpoints to provide access to these replicas.
%\end{definition}

\begin{definition}[Containment relation among fragments]
A fragment F1 is contained in a fragment F2, if all the possible triples that may belong to F1
may also belong to F2.
\end{definition}

\begin{definition}[Divergence]
Divergence is the different between a fragment and its replicas.
\end{definition}
This difference may be measured in different ways, some examples are elapsed time 
and in number of performed operations.

%\begin{definition}[SPARQL federation]
%SPARQL federations are constituted by a large number of participants that provide
%SPARQL endpoints giving access to their resources.
%\end{definition}

\begin{definition}[Fragment relevance]
A data fragment $f$ is relevant for answering a query $Q$ if the query that defines $f$ and 
$Q$ have some compatible triple patterns, and it is denoted as $relevant(f, Q)$. 
Two triple patterns $t1$ and $t2$ are compatible 
if there exists at least one possible binding of values to variables in $t1$, $t2$ such that the 
resulting triples after applying the binding are equal.
\end{definition}

\spnewtheorem*{SourceSelection}{Source Selection Problem (SSP)}{\bfseries}{\rmfamily}
\begin{SourceSelection}
Given a set of SPARQL endpoints $E$, 
a subset $P$ of $E$ that correspond to the public endpoints, 
the set of fragments contained in each endpoint 
as a function $fragments \, : \, Endpoint \rightarrow \, set \, of \, Fragment$,
a containment relation among endpoints for each $frag \in fragments(e_i) \land frag \in fragments(e_j)$,
 $e_i \,\subseteq_{frag} \,e_j$, the age or divergence of each fragment $frag$ replicated 
 in $e_i$ from $e_j$, $e_i \,div_{frag} \,e_j$, a query $Q$ and an age limit or divergence 
 tolerance $dt$, find a map D, such that for each triple pattern $t$ in $Q$ $D(t) \subseteq(E)$ and:
\begin{itemize}
  \item answer($Q$, $D$) $=_{dt}$ answer($Q$, $F$), where F is a map such that for each triple pattern $t$ in $Q$ has value $E$.
  \item $(\forall t : t \in triplePatterns(Q) : (\forall e, f : e \in E – D(t), f \in fragments(e),$ \\$relevant(f, Q) : (\exists e': e' \in D(t) : e \subseteq_f e')))$
  \item $D(t)$ contains as few public endpoints as possible.
  \item $\bigcup_{t \,in \,triplePatterns(Q)} D(t)$ is a minimal set that satisfies these conditions
\end{itemize}
\end{SourceSelection}

%\fedra enables SPARQL federated engines to take advantage of data replication during  query processing.
Figure~(\ref{datasetD}) illustrates an opportunistic federation with fragments from datasets E and F.
D fragments ($D1$, $D2$, and $D3$) are defined by views $V1$ and $V2$, and F fragments ($F1$ and $F2$) are defined by views $V3$ and $V4$. $V1$ defines a fragment that consists of the whole dataset, while $V2$, $V3$, and $V4$ define fragments that consist of a subset of the corresponding dataset.
There is an explicit containment relationship among these fragments, for example fragment $D1$ data in endpoint $E3$ comes from endpoint fragment $D$ in endpoint $E1$, \emph{i.e.},  $D \sqsubseteq_{V1} D1$.
There is also an implicit containment relationship, for example fragment D2 data is contained in fragment D1 data. 
The public SPARQL endpoint $E1$ provides access to $D$, while local SPARQL endpoints $E3$, $E4$, and 
$E5$  access $D1$, $D2$, and $D3$, respectively. 
In addition,  a replica endpoint may provide access to fragments from different datasets like $E5$ that 
combines data from $D3$ and $F1$.
At the moment T+2, fragments $D3$ and $F1$ are up to 
date because their last included update was at time T+2, while fragments $D1$ and $F2$ have age 1, because their last included update was at time T+1, and
fragment $D2$ has age 2 because its last included update was at time T.
All the SPARQL endpoints can contribute during federated query processing. Therefore,  
workload on public endpoints can be reduced, and replica SPARQL endpoints can replace the public ones 
whenever they are not available. 

\newsavebox{\tempbox}
\begin{figure}
\begin{center}
\subfloat[Fragments of Datasets D and F]
{\label{datasetD}
\begin{tikzpicture}%[->,>=stealth',shorten >=1pt,auto,node distance=1.5cm,
                    %thick,main node/.style={circle,draw,font=\sffamily\tiny\bfseries}]
  \tikzstyle{every node}=[font=\tiny];
  %\draw(E2) (0,0) rectangle (2, 1);
  %\draw(E3) (5,0) rectangle (7, 1);
  %\draw(E1) (2.5,5) rectangle (4.5, 6);
  \node[draw, text width=1.8cm, text height=0.5cm,align=center] (E2) at (0,-0.3) {\small s1 p1 o1 .\\o1 p2 o2.\\s2 p1 o3.\\o3 p3 o4.\\o1 p2 o7};
  \node at (2.3,4.3) {\small E1$^{*}$=\{D\}} ;
  \node at (0,1.2) {\small E3=\{D1\}} ;
  \node at (2.35,1.2) {\small E4=\{D2\}} ;
  \node at (5.8,1.2) {\small E5=\{D3,F1\}} ;
  \node at (1.6,1.6) {\small V1} ;
  \node at (2.55,1.6) {\small V2} ;
  \node at (3.6,1.6) {\small V1} ;
  \node[draw, text width=1.8cm, text height=0.15cm,align=center] (E1) at (2.3,3) {\small s1 p1 o1 .\\o1 p2 o2.\\s2 p1 o3.\\o3 p3 o4.\\o1 p2 o7.\\o3 p3 o8};
  \node[draw, text width=1.8cm, text height=1.75cm,align=center] (E3) at (2.3,-0.3) {\small s1 p1 o1 .\\o1 p2 o2};
  \node[draw, text width=1.8cm, text height=0.2cm,align=center] (E4) at (4.6,-0.3) {\small s1 p1 o1 .\\o1 p2 o2.\\s2 p1 o3.\\o3 p3 o4.\\o1 p2 o7.\\o3 p3 o8};
  \node[draw, text width=1.8cm, text height=1.75cm,align=center] (E5) at (6.9,-0.3) {\small o2 p4 o5 . o2 p4 o6};
  \node[draw, text width=1.8cm, text height=1cm,align=center] (E6) at (8.05,3) {\small o1 p4 o3 .\\o2 p4 o5.\\o2 p4 o6.\\o4 p4 o3};
  \node[draw, text width=1.8cm, text height=1.75cm,align=center] (E7) at (9.2,-0.3) {\small o1 p4 o3 .\\o4 p4 o3};  
  \node[draw, fit=(E4) (E5)] {};
  \node[draw, fit=(E7)] {};
  \node[draw, fit=(E2)] {};
  \node[draw, fit=(E3)] {};
    \node at (8,4.3) {\small E2$^{*}$=\{F\}} ;
  \node at (9.5,1.2) {\small E6=\{F2\}} ;
  \node at (7.2,1.6) {\small V3} ;
  \node at (8.9,1.6) {\small V4} ;
   \draw[->] (E1) -- (E2);
  \draw[->] (E1) -- (E3);
  \draw[->] (E1) -- (E4);
  \draw[->] (E6) -- (E5);
  \draw[->] (E6) -- (E7);
\end{tikzpicture}
}
%\subfloat[Fragments of Dataset F]
%{\label{datasetF}
%\begin{tikzpicture}%[->,>=stealth',shorten >=1pt,auto,node distance=1.5cm,
                    %thick,main node/.style={circle,draw,font=\sffamily\tiny\bfseries}]
%  \tikzstyle{every node}=[font=\tiny];
  %\draw(E2) (0,0) rectangle (2, 1);
  %\draw(E3) (5,0) rectangle (7, 1);
  %\draw(E1) (2.5,5) rectangle (4.5, 6);
%  \node at (1.5,4.1) {\small F} ;
%  \node at (0,1.1) {\small F1} ;
%  \node at (3.2,1.1) {\small F2} ;
%  \node at (0.6,1.8) {\small V3} ;
%  \node at (2.5,1.8) {\small V4} ;
 % \node[draw, text width=1.8cm, text height=1.3cm,align=center] (E2) at (0,0) {\small o2 p4 o5 . o2 p4 o6};
 % \node[draw, text width=1.8cm, text height=0.5cm,align=center] (E1) at (1.5,3) {\small o1 p4 o5 .\\o2 p4 o5.\\o2 p4 o6.\\o4 p4 o6};
  %\node[draw, text width=1.8cm, text height=1.3cm,align=center] (E3) at (3,0) {\small o1 p4 o5 .\\o2 p4 o5};
   %\draw[->] (E1) -- (E2);
  %\draw[->] (E1) -- (E3);
%\end{tikzpicture}
%%}
\\
\subfloat[D last operations]%
{\label{insertionsInD}%
\small
%\vbox  to \ht\tempbox{%
%\vfil
\begin{tabular}{|c|c|c|}
 \hline
 Date & Type & Operation \\
 \hline
 T & I & s1 p1 o1 \\
    & I & o1 p2 o2 \\
    & I & s2 p1 o3 \\
    & I & o3 p3 o4 \\
    \hline
T+1 & I & o1 p2 o7\\
\hline
T+2 & I & o3 p3 o8\\
 \hline
\end{tabular}
%\vfil
%}
}
\subfloat[Views definitions]
{\label{fragmentsDescription}
\begin{tabular}{|c|c|}
 \hline
View& Definition \\
 \hline
V1 & CONSTRUCT WHERE \{ ?x p1 ?y . ?y ?p ?z \} \\
      \hline
V2 & CONSTRUCT WHERE \{ ?x p1 ?y . ?y p2 ?z \} \\
\hline
V3 & CONSTRUCT WHERE \{ o2 p4 ?x \} \\
\hline
V4 & CONSTRUCT WHERE \{ ?x p4 o3 \} \\
 \hline
\end{tabular}
}\\
\subfloat[Fragments age]
{\label{endpointsContents}
\begin{tabular}{|c|c|c|c|c|c|c|c|}
 \hline
 Endpoint[Fragment] & E3[D1] & E4[D2] & E5[D3] & E5[F1] & E6[F2] \\
 \hline
 Age & 1 & 2 & 0 & 0 & 1\\
 \hline
\end{tabular}
}
\caption{Example of Opportunistic Federation, for each endpoint on the top it is indicated the contained fragments, endpoints marked with $^{*}$ are public endpoints. Updates for the last three time unit, fragments view definitions and fragment age.}
\label{fig:partialRep}
\end{center}
\end{figure}

\begin{minipage}{0.9\linewidth}
 \begin{lstlisting}[basicstyle=\small\sffamily,language=sparql,caption={Query Q1},numbers=none,frame=none,columns=fixed,label=query,extendedchars=true,breaklines=true,showstringspaces=false]
SELECT DISTINCT ?s ?o ?r
WHERE {
  ?s p1 ?o .
  ?o p4 ?r
}
\end{lstlisting}
\end{minipage}

Consider the query given in Listing~\ref{query} with triple patterns 
k1=($?s \,p1 \,?o$) and k2=($?o \,p4 \,?r$), and an age limit of 1.
A solution to the SSP is \{ (k1, \{ E5 \}), (k2, \{ E5, E6\})\}.
For k1 endpoints E1, E3, E4, and E5 are relevant, and from these E1, E3 and E5 are
equivalent, i.e., there is a containment in both senses, and E4 is strictly contained
in them. Then, choosing E5 for k1 satisfies that all the other endpoints that provide
relevant data for k1 are ``covered'' by E5. For k2 endpoints E2, E5 and E6 are relevant,
and from these E2 provides as much information as both E5 and E6 combined, but E2
is a public endpoint, then E5 and E6 should be selected.
Notice that the choice of E5 for k1 is necessary to reduce the total number of 
selected sources, since it is also relevant for k2.
Selected endpoints, $E5$ and $E6$, satisfy the age limit of 1.

\subsection{Source Selection Algorithm}

\begin{algorithm}[H] 
\scriptsize
%\algsetup{linenosize=\footnotesize}
 \caption{Source Selection algorithm}
 \label{sourceselectionalgorithm}
\begin{algorithmic}[1]
\Require $Q$: SPARQL Query
\Require $E$: set of Endpoints
\Require $fragments\;:\;Endpoint \rightarrow set \; of \; fragment$ \Comment {fragments offered by each endpoint}
%\Require $origin\;:\;Endpoint \;\times\; view \rightarrow set \; of\; Endpoint$ \Comment {endpoints from which the content of each view is taken}
\Require $P \; : set \, of \, Endpoint$ \Comment {endpoints that should not be selected}
\Require $\subseteq_{frag} \; : \; Endpoint \times Endpoint$ \Comment {containment relation given by fragments}
\Require $div_{frag} \; : \; Endpoint \times Endpoint \rightarrow Integer$ \Comment {Fragment age}
\Require $dt \; : \; Integer$ \Comment{Age limit}
\Ensure $D$: Dictionary From Triple Pattern to List of Endpoints.
\Function{sourceSelection}{$Q$,$E$, $fragments$, $P$, $\subseteq_{frag}$, $div_{frag}$, $dt$}
\State triplePatterns $\leftarrow$ get triple patterns in Q
\For  {each $k \in triplePatterns$} 
   \State G(k) $\leftarrow$ get candidates c from E, such that c div$_frag$ frag.origin $\leq$ dt  
   \State split G(k) into set of endpoints that provide the same data fragment
   \State simplify G(k) using containment among views
   \State for the sets in G(k) that are not singleton of publicEndpoint, remove publicEndpoint.
\EndFor
%\State \Comment {S is obtained from the triple patterns, and C from the endpoints}
\State (S, C) $\leftarrow$ get instance of minimal set covering problem using G
\State C' $\leftarrow$ minimalSetCovering(S, C)
\For  {each $k \in domain(G)$} 
   \State G(k) $\leftarrow$ filter G(k) according to C'
   \State D(k) $\leftarrow$ for each set in G(k) include one of its elements
\EndFor
\State \Return D
\EndFunction
\end{algorithmic}
\end{algorithm}

Algorithm~\ref{sourceselectionalgorithm} presents the source selection pseudocode.
First, this algorithm pre-select for each triple pattern in the query the sources that can be used to
evaluate it, and satisfies the divergence threshold provided by the user. 
The capability of a source to provide data for a triple pattern can 
be implemented using an {\tt ASK} query for dynamic data, or relying on fragment definition
for more stable data.
In the example given above, G = \{ (k1, \{E1, E3, E4, E5 \}), (k2, \{ E2, E5, E6\} \}.
Second, these pre-selected sources are grouped according to the 
containment relation between them, some sources can provide the same fragment of data or the data
provide by one source may be contained in the data provided by another source.
At the end of the first for loop (lines3-8), a list whose elements are list of equivalent endpoints is produced for each
triple pattern in the query. This means that they offer the same fragment given their view definition, 
then during execution only one of them needs to be contacted. And different 
elements of this resulting list correspond to different fragments that should be considered 
in order to obtain an answer as complete
as possible, modulo the considered endpoints and the allowed divergence threshold.
In the example given above, E1, E3 and E5 are grouped together since they provide access
to the same fragment, then at the end of this second step
G = \{ (k1, \{ \{E1, E3, E5\}, \{ E4\} \}), (k2, \{ \{E2, E5\}, \{ E2, E6\}\} \}.
Third, these pre-selected sources are simplified according to the containment relation among them.
Then the source \{ E4 \} is removed from G(k1), because it provides a fragment already covered by \{E1, E3, E5\}.
Fourth, the public endpoints are removed when possible. In the example, endpoint E2 is removed since its contents
can be accessed using E5 and E6.
Fifth, a general selection takes place, considering the pre-selected sources for 
each triple pattern in the query. This last part can be reduce to the well-known
set covering problem, and an existing heuristic like the one given 
in~\cite{DBLP:books/daglib/0017733} may be used to perform the procedure 
indicated in line 10. 
To illustrate this reduction, in the example $G = \{ (k1, \{ \{E1, E3, E5\}\}), (k2, \{ \{ E5\}, \{ E6\}\} )\}$, corresponds
to $S = \{ k1_1, k2_1, k2_2 \}$ and $C = \{ E1, E3, E5, E6 \}$, $E1 = E3 = \{ k1_1 \}, E5 = \{ k1_1, k2_1 \}$, $E6 = \{ k2_2 \}$. Here $ki_j$ represent the different fragments of $ki$ that should be covered to produce a 
complete answer. In the case of $k1$, only one fragment should be covered, and in the case of $k2$
two different fragments should be covered $k2_1$ and $k2_2$.
Then a minimal set covering solution $C' = \{ E5, E6 \}$ covers all set $S$, and has minimal size.
This solution is used in line 12 to filter $G(k)$ for each $k$, then $G = \{ (k1, \{ \{E5\}\}), (k2, \{ \{ E5\}, \{ E6\}\} \}$.
In this case, each element of $G(k)$ is a singleton, but otherwise a last step may be performed to choose among
endpoints that provide the same fragment and ensure a better query processing by existing federated query 
engines.
Nevertheless, these alternative sources could be used to speed up query processing, 
for example by getting a part of the answer from each endpoint and combining them.

\begin{proposition}
Let n be the number of triple pattern in the query, s be the number of available sources, and v be the
maximal number of views that describe a source,
the time complexity of Algorithm~\ref{sourceselectionalgorithm} is $O(n \times s \times Max(s, v))$
\end{proposition}

%1/ \fedra takes advantage of opportunistic replication than FEDX at the price
%bounded stales values
%2/ \fedra improve availability and scalability (as in LDF ??)

% We should explain how this descriptions are updated over time..
\section{Experiments}
\label{sec:experiments}

The aim of this section is to give evidence of \fedra benefits over existing approaches.
We compare \fedra to DAW source selection algorithm as it is the closest approach for federations
of SPARQL endpoints. The selected sources are used to annotate SPARQL queries with the SERVICE clause.
Execution of the annotated queries is performed using FedX, additionally direct FedX execution is used as
baseline.

% In order to validate \fedra approach, the following research questions should be 
% answered by the reported experiments:
% \begin{inparaenum}[\itshape RQ1\upshape)]
% \item does \fedra select less sources than existing source selection techniques?,
% \item if so, can we characterize the cases where larger and smaller reduction can 
%           be achieved?,
% \item more importantly, does this reduction bring some recall reduction?,
% \item is the reduction of selected public endpoints significative?,
% \item how much is the cost of \fedra source selection algorithm in practice?,
% \item \fedra can select sources that satisfy a user threshold of divergence. 
%          How does the quality of the divergence metric, and the increase of the 
%          threshold affect the completeness and correctness of the answer?, and
% \item does \fedra contribute to improve the data availability?
% \end{inparaenum}

 \begin{table}[ht]
 \begin{center}
\caption{Queries caracteristics}
\label{table:queries}
{\scriptsize
\begin{tabular}{|c|c|c|c|c|c|c|c|c|}
\hline
Queries & \# Answers & \# Triple Patterns & Distinct & Union & Filter & Optional & Regex & Bound\\
\hline
q1 & 3 & 1 & 1 & 0 & 0 & 0 & 0 & 0\\
q2 & 1 & 1 & 1 & 0 & 0 & 0 & 0 & 0\\
q3 & 720 & 5 & 0 & 0 & 0 & 1 & 0 & 0\\
q4 & 39 & 5 & 0 & 2 & 0 & 0 & 0 & 0\\
q5 & 9310 & 4 & 0 & 0 & 1 & 0 & 0 & 0\\
\hline
q6 & 5 & 1 & 0 & 0 & 1 & 0 & 0 & 0\\
q7 & 8 & 9 & 1 & 8 & 0 & 0 & 0 & 0\\
q8 & 5 & 1 & 0 & 0 & 1 & 0 & 0 & 0\\
q9 & 0 & 3 & 0 & 0 & 1 & 0 & 0 & 0\\
q10 & 1814176 & 2 & 0 & 0 & 1 & 0 & 1 & 0\\
\hline
q11 & 0 & 2 & 0 & 0 & 2 & 1 & 0 & 1\\
q12 & 12840 & 4 & 0 & 0 & 2 & 2 & 0 & 0\\
q13 & 2 & 3 & 1 & 0 & 1 & 0 & 0 & 0\\
q14 & 1 & 1 & 1 & 0 & 1 & 0 & 0 & 0\\
q15 & 4 & 3 & 0 & 2 & 1 & 0 & 0 & 0\\
\hline
q16 & 61089 & 2 & 1 & 0 & 1 & 1 & 0 & 0\\
q17 & 6 & 2 & 0 & 1 & 2 & 0 & 0 & 0\\
q18 & 32 & 1 & 0 & 0 & 0 & 0 & 0 & 0\\
q19 & 163 & 2 & 0 & 0 & 0 & 0 & 0 & 0\\
q20 & 312 & 3 & 0 & 0 & 0 & 0 & 0 & 0\\
\hline
\end{tabular}}
\end{center}
\end{table}

\noindent
{\bf Dataset, Queries and Federations Benchmark:} we used the last three DBpedia Live dataset dumps 
available~\footnote{\url{http://live.dbpedia.org/dumps/dbpedia_2013_07_18.nt.bz2}}~\footnote{\url{http://live.dbpedia.org/dumps/dbpedia_2013_06_17.nt.bz2}}\footnote{\url{http://live.dbpedia.org/dumps/dbpedia_2013_05_17.nt.bz2}}, 
%the live updates from a 14 days period (July 18 - August 1, 
%2013)~\footnote{http://live.dbpedia.org/liveupdates/2013/},
and queries from the DBpedia SPARQL Benchmark~\cite{DBLP:conf/semweb/MorseyLAN11}. 
Table~\ref{table:queries} summarize the characteristics of the 
 queries.
600 Fragments of this dataset were defined using random basic graph pattern 
queries that concern the benchmark queries, having 100 fragments of each size between 1 and 6.
The fragments were randomly distributed in 100 endpoints, each fragment may be assigned to 0-3 endpoints.

%In order to study the impact of data replication we consider opportunistic federations where 
%data consumers of this dataset have decided to share local SPARQL endpoints that expose data fragments 
%that they use for their own purposes with other users. 

In order to characterize \fedra behavior, we set up federations that comprise a varying number 
of participants: 10, 25, 50 and 100.

Virtuoso\footnote{\url{http://virtuoso.openlinksw.com/}, November 2013.} endpoints are used, and 
timeouts were set up to 1,800 secs. and 100,000 tuples. 
Listing~(\ref{viewB}) presents a query that defines a fragment~\footnote{All the fragments 
definitions and federations configurations are available at  the project website: \url{https://sites.google.com/site/fedrasourceselection/}}.

\noindent
{\bf Divergence among replicated fragments:} To study the impact of data updates, and divergence 
 over the staleness of query answers, federations members have included fragments taken from the
 three considered epochs of the dataset.

\def\lstlistingname{Listing}

% \begin{minipage}{0.55\linewidth}
%  \begin{lstlisting}[basicstyle=\small\sffamily,language=sparql,caption={DBpedia fragment 1},numbers=none,frame=none,columns=fixed,label=viewA,extendedchars=true,breaklines=true,showstringspaces=false]
% CONSTRUCT WHERE {
%   ?x1  rdf:type  ?x3 .
% }
% \end{lstlisting}
% \end{minipage}
\begin{minipage}{0.65\linewidth}
 \begin{lstlisting}[basicstyle=\small\sffamily,language=sparql,caption={DBpedia fragment 2},numbers=none,frame=none,columns=fixed,label=viewB,extendedchars=true,breaklines=true,showstringspaces=false]
CONSTRUCT WHERE {
  ?x1  rdf:type  ?x3 .
  ?x1  dbpedia:nearestCity  ?x4 .
  ?x1  dbpedia:iucnCategory  ?x5 .
}
\end{lstlisting}
\end{minipage}

 \noindent
{\bf Implementations:} \fedra is
implemented using Java 1.7 and the Jena 2.11.0 library~\footnote{\url{http://jena.apache.org/}, November 2013.}. 
\fedra produces SPARQL 1.1 queries where each triple pattern is annotated with a service clause that indicates 
where it will be executed. We have done a likewise reference implementation of 
DAW~\cite{DBLP:conf/semweb/SaleemNPDH13}, as its code is not available for comparison.
These queries are posed to FedX3.0\footnote{\url{http://www.fluidops.com/fedx/}, November 2013.}. FedX is an state-of-the-art 
federated engine that process both 1.0 and 1.1 SPARQL queries, this characteristic is crucial to show 
that \fedra benefits the engine with respect to its source selection strategy.

\noindent
{\bf Evaluation Metrics:}
\begin{inparaenum}[\itshape i\upshape)]
 \item {\it Number of Selected Public Sources (NSPS):} corresponds to the sum of the 
 number of times the public endpoint has been selected per triple pattern.
 \item {\it Number of Selected Sources (NSS):} corresponds to the sum of the number of 
 sources that has been selected per triple pattern.
 \item {\it Execution Time (ET):} corresponds to elapsed time since the query is posed by the user and the answers are
  completely produced. It is detailed in source selection time (SST), and query execution by the underlying engine (ETUE). 
 Time is expressed in seconds (secs.). A timeout of 600 secs. has been enforced. Time was measured using 
 System.currentTimeMillis() provided by Java and divided by 1000.
 \item {\it Completeness (C):} corresponds to the size of the bag intersection between the obtained answers and the expected 
 answers divided by the number of expected answers; where the expected answers are the ones obtained from an 
 endpoint containing the whole dataset.
 \item {\it Staleness (S):} corresponds to the size of the bag difference between the obtained answers and all the 
  occurrences of the expected answers, divided by the number of obtained answers.
 \item {\it Divergence (Div):} corresponds to the distance between a replicated fragment and the same 
 fragment in the public endpoint. %The burden imposed by computing this metric by participant should 
 %remain low to keep the approach attractive to participants. 
 A $\Delta$-distance is
 used, it  corresponds to the elapsed time since the last updated was included.
%   \item {\it Availability (Av):} it corresponds to the assessed reliability of DBpedia Live SPARQL endpoint 
%    by the SPARQL Endpoints monitoring tool SPARQLES\footnote{\url{http://sparqles.okfn.org/}}.
%  \item {\it Execution Time Under Availability (ETUA):} it corresponds to the adjusted query execution time, considering the 
%  public SPARQL endpoint availability. Similary to~\cite{verborgh_ldow_2014}, we compute it as ET + (1-Av)*(900/2).
\end{inparaenum}

\subsection{Impact of the Age Limit over the Results Completeness and Staleness}

% % Mention the reduction of the public endpoints
% \begin{figure}
% \begin{center}
% \subfloat[Number of selected sources]{
% \label{fig:numberselectedsources}
% \includegraphics[scale=0.46]{numberSelectedSources}
% }\hspace*{-0.2cm}
% \subfloat[Execution time]{
% \label{fig:sourceselectiontime}
% \includegraphics[scale=0.46]{sourceSelectionTime}
% }
% \caption{Comparison of the number of Number of Selected Sources (NSS) for \fedra, DAW and
% FedX; and comparison of the source selection time (SST) and execution time for queries with and without SERVICE annotation (ET).}
% \label{fig:results}
% \end{center}
% \end{figure}

  \begin{table}[!htbp]
 \begin{center}
\caption{Impact of the age limit (0, 1 and 2 months) over the \fedra source selection and subsequent execution by FedX.}
{
\scriptsize
\begin{tabular}{|c|r|r|r|r|r|r|r|r|}
 \hline
Query & \multicolumn{1}{c|}{Div} & NSS & NSPS & \multicolumn{1}{c|}{SST} & \multicolumn{1}{c|}{ETUE} & \# Answers & \multicolumn{1}{c|}{C} & \multicolumn{1}{c|}{S}\\
 \hline
q1 & 0 months & 1 & 1 & 0,7670 & 0,9400 & 3 & 1,0000 & 0,0000\\
 & 1 month & 1 & 0 & 0,9840 & 1,3000 & 3 & 1,0000 & 0,0000\\
 & 2 months & 1 & 0 & 0,5280 & 2,1300 & 3 & 1,0000 & 0,0000\\
  \hline
q2 & 0 months & 1 & 1 & 0,3090 & 0,9900 & 1 & 1,0000 & 0,0000\\
 & 1 month & 1 & 1 & 0,9090 & 1,7700 & 1 & 1,0000 & 0,0000\\
 & 2 months & 2 & 0 & 0,3080 & 3,2400 & 1 & 1,0000 & 0,0000\\
  \hline
q3 & 0 months & 5 & 3 & 2,8300 & 39,0500 & 144 & 0,2000 & 0,0000\\
 & 1 month & 5 & 2 & 7,6080 & 64,9000 & 144 & 0,2000 & 0,0000\\
 & 2 months & 5 & 0 & 3,6560 & 1,6700 & 0 & 0,0000 & 0,0000\\
  \hline
q4 & 0 months & 5 & 5 & 1,1090 & 1,5900 & 39 & 1,0000 & 0,0000\\
 & 1 month & 5 & 2 & 3,3610 & 0,8200 & 39 & 1,0000 & 0,0000\\
 & 2 months & 5 & 2 & 0,9990 & 13,9000 & 39 & 1,0000 & 0,0000\\
  \hline
q5 & 0 months & 12 & 3 & 2,0210 & 5,7500 & 83790 & 1,0000 & 0,0000\\
 & 1 month & 17 & 2 & 9,0570 & 300,0000 & 0 & 0,0000 & 0,0000\\
 & 2 months & 18 & 0 & 2,7170 & 47,2500 & 139649 & 1,0000 & 0,0000\\
  \hline
q6 & 0 months & 1 & 0 & 1,4240 & 0,9400 & 5 & 1,0000 & 0,0000\\
 & 1 month & 1 & 0 & 2,0030 & 0,8800 & 5 & 1,0000 & 0,0000\\
 & 2 months & 1 & 0 & 0,9380 & 1,9600 & 5 & 1,0000 & 0,0000\\
  \hline
q7 & 0 months & 12 & 3 & 3,0350 & 0,9700 & 8 & 1,0000 & 0,0000\\
 & 1 month & 15 & 1 & 7,4240 & 0,9300 & 8 & 1,0000 & 0,0000\\
 & 2 months & 14 & 0 & 4,0070 & 5,7700 & 8 & 1,0000 & 0,0000\\
  \hline
q8 & 0 months & 1 & 1 & 1,3310 & 0,9600 & 5 & 1,0000 & 0,0000\\
 & 1 month & 1 & 0 & 2,9670 & 0,8900 & 5 & 1,0000 & 0,0000\\
 & 2 months & 1 & 0 & 0,9030 & 2,5700 & 5 & 1,0000 & 0,0000\\
  \hline
q9 & 0 months & 3 & 3 & 1,1620 & 0,9300 & 0 & 1,0000 & 0,0000\\
 & 1 month & 3 & 1 & 2,2580 & 0,9900 & 576 & 1,0000 & 1,0000\\
 & 2 months & 3 & 1 & 1,5160 & 2,1900 & 576 & 1,0000 & 1,0000\\
  \hline
q10 & 0 months & 10 & 2 & 1,2430 & 300,0000 & 123771 & 0,0000 & 0,0000\\
 & 1 month & 13 & 2 & 3,3350 & 300,0000 & 165026 & 0,0000 & 0,0048\\
 & 2 months & 15 & 0 & 1,5110 & 300,0000 & 65772 & 0,0000 & 0,0000\\
  \hline
q11 & 0 months & 2 & 2 & 0,8010 & 0,9200 & 6 & 1,0000 & 1,0000\\
 & 1 month & 2 & 1 & 1,0720 & 1,2100 & 6 & 1,0000 & 1,0000\\
 & 2 months & 2 & 0 & 0,7360 & 1,8100 & 6 & 1,0000 & 1,0000\\
  \hline
q12 & 0 months & 4 & 2 & 1,5140 & 300,0000 & 0 & 0,0000 & 0,0000\\
 & 1 month & 4 & 1 & 4,2710 & 300,0000 & 0 & 0,0000 & 0,0000\\
 & 2 months & 4 & 0 & 2,7910 & 300,0000 & 0 & 0,0000 & 0,0000\\
  \hline
q13 & 0 months & 3 & 2 & 1,9270 & 1,3000 & 2 & 1,0000 & 0,0000\\
 & 1 month & 3 & 1 & 3,7320 & 1,5200 & 2 & 1,0000 & 0,0000\\
 & 2 months & 3 & 0 & 1,9500 & 2,1600 & 2 & 1,0000 & 0,0000\\
  \hline
q14 & 0 months & 1 & 0 & 0,6700 & 0,9200 & 1 & 1,0000 & 0,0000\\
 & 1 month & 1 & 0 & 0,7020 & 2,0900 & 1 & 1,0000 & 0,0000\\
 & 2 months & 1 & 0 & 0,6170 & 2,1200 & 1 & 1,0000 & 0,0000\\
  \hline
q15 & 0 months & 1 & 0 & 1,3200 & 0,9600 & 4 & 1,0000 & 0,0000\\
 & 1 month & 1 & 0 & 3,6130 & 1,3000 & 4 & 1,0000 & 0,0000\\
 & 2 months & 1 & 0 & 2,3030 & 1,2700 & 4 & 1,0000 & 0,0000\\
  \hline
q16 & 0 months & 2 & 2 & 1,6320 & 278,3700 & 21642 & 0,3500 & 0,0004\\
 & 1 month & 2 & 1 & 1,3620 & 300,0000 & 14546 & 0,2300 & 0,0327\\
 & 2 months & 2 & 0 & 1,6940 & 300,0000 & 14542 & 0,2300 & 0,0327\\
  \hline
q17 & 0 months & 1 & 0 & 1,9510 & 1,3000 & 0 & 0,0000 & 0,0000\\
 & 1 month & 1 & 0 & 2,1440 & 1,4300 & 0 & 0,0000 & 0,0000\\
 & 2 months & 1 & 0 & 2,0670 & 2,1400 & 0 & 0,0000 & 0,0000\\
  \hline
q18 & 0 months & 1 & 1 & 0,9960 & 1,6000 & 32 & 1,0000 & 0,0000\\
 & 1 month & 1 & 0 & 0,7250 & 1,2200 & 32 & 1,0000 & 0,0000\\
 & 2 months & 1 & 0 & 0,6580 & 1,9000 & 32 & 1,0000 & 0,0000\\
  \hline
q19 & 0 months & 2 & 2 & 1,4400 & 1,3300 & 163 & 1,0000 & 0,0000\\
 & 1 month & 2 & 1 & 1,5190 & 0,8000 & 163 & 1,0000 & 0,0000\\
 & 2 months & 2 & 0 & 1,4270 & 2,4400 & 163 & 1,0000 & 0,0000\\
 \hline
 q20 & 0 months & 3 & 3 & 1,9240 & 26,1400 & 24 & 0,0700 & 0,0000\\
 & 1 month & 3 & 2 & 2,4330 & 95,7200 & 24 & 0,0700 & 0,0000\\
 & 2 months & 3 & 0 & 2,1810 & 94,3100 & 24 & 0,0700 & 0,0000\\
 \hline
\end{tabular}
}
\label{table:onehundredfederation}
\end{center}
\end{table}

Table~\ref{table:onehundredfederation} shows the source selection and execution results over 
a 100 endpoints federation for three different age limit.
As the age limits increases up to two months, the number of stale values in the answer remain low, and the
answer completeness remain high. Additionally, increassing the age limit allows to consider more
appropiate replicas to execute the queries and in consequence there is a reduction in
number of times the public endpoint
is selected.

\subsection{Reduction of the Number of Selected Sources}

  \begin{table}[!htbp]
 \begin{center}
\caption{\fedra Steadiness of the Number of Selected Sources (NSS), the results correspond to the average of five executions over random federations}
{
\scriptsize
\begin{tabular}{|c|c|r|r|r|r|r|r|r|r|r|}
 \hline
Query&Approach & \multicolumn{3}{c|}{10 endpoints Federation} & \multicolumn{3}{c|}{25 endpoints Federation} & \multicolumn{3}{c|}{50 endpoints Federation}\\
\cline{3-11}
&& NSS & NSPS & \multicolumn{1}{c|}{SST} & NSS & NSPS & \multicolumn{1}{c|}{SST} & NSS & NSPS & \multicolumn{1}{c|}{SST}\\
\hline
q1 & FedX & 10.8 & 1.0 & 1.34 & 24.2 & 1.0 & 1.32 & 47.4 & 1.0 & 1.62\\
 & DAW & 3.6 & 0.0 & 0.30 & 4.4 & 0 & 0.28 & 11.0 & 0.0 & 0.35\\
 & \fedra & 1.0 & 1.0 & 0.30 & 1.0 & 0.0 & 0.32 & 1.0 & 1.0 & 0.41\\
  \hline
q2 & FedX & 21.6 & 1.0 & 1.55 & 24.2 & 1.0 & 1.61 & 47.4 & 1.0 & 1.72\\
 & DAW & 1.0 & 0.6 & 0.27 & 1.0 & 0.4 & 0.23 & 1.0 & 0.0 & 0.23\\
 & \fedra & 1.0 & 1.0 & 0.27 & 1.0 & 1.0 & 0.27 & 1.0 & 1.0 & 0.31\\
  \hline
q3 & FedX & 21.6 & 5.0 & 1.37 & 121.8 & 5.0 & 1.42 & 238.6 & 5.0 & 1.94\\
 & DAW & 34.4 & 4.0 & 0.64 & 76.4 & 3.2 & 0.85 & 143.6 & 2.4 & 1.96\\
 & \fedra & 5.0 & 5.0 & 0.62 & 5.0 & 4.0 & 1.30 & 5.0 & 4.0 & 2.12\\
  \hline
q4 & FedX & 43.2 & 5.0 & 1.52 & 121.0 & 5.0 & 1.54 & 237.0 & 5.0 & 1.74\\
 & DAW & 12.8 & 2.0 & 0.44 & 15.2 & 2 & 0.40 & 35.0 & 2.0 & 0.68\\
 & \fedra & 5.0 & 5.0 & 0.43 & 5.0 & 2.0 & 0.48 & 5.0 & 5.0 & 0.84\\
  \hline
q5 & FedX & 32.4 & 4.0 & 1.35 & 98.0 & 4.0 & 1.53 & 192.0 & 4.0 & 1.64\\
 & DAW & 25.4 & 2.4 & 0.55 & 57.0 & 2.2 & 0.81 & 108.4 & 2.0 & 1.22\\
 & \fedra & 6.6 & 3.0 & 0.54 & 7.8 & 3.0 & 0.84 & 12.4 & 4.0 & 1.45\\
  \hline
q6 & FedX & 10.8 & 1.0 & 1.32 & 24.2 & 1.0 & 1.53 & 47.4 & 1.0 & 1.36\\
 & DAW & 10.0 & 1.0 & 0.48 & 23.0 & 1 & 0.65 & 43.2 & 1.0 & 1.66\\
 & \fedra & 1.0 & 0.0 & 0.34 & 1.0 & 0.0 & 0.43 & 1.0 & 0.0 & 0.69\\
  \hline
q7 & FedX & 10.8 & 4.0 & 1.35 & 96.8 & 4.0 & 1.51 & 189.6 & 4.0 & 1.93\\
 & DAW & 16.6 & 0.0 & 0.78 & 29.0 & 0 & 1.23 & 54.8 & 0.0 & 2.31\\
 & \fedra & 5.0 & 4.0 & 0.79 & 8.0 & 2.0 & 1.58 & 10.2 & 3.0 & 2.33\\
  \hline
q8 & FedX & 21.6 & 1.0 & 1.45 & 24.2 & 1.0 & 1.51 & 47.4 & 1.0 & 1.50\\
 & DAW & 4.6 & 0.0 & 0.38 & 11.8 & 0 & 0.38 & 20.8 & 0.0 & 0.58\\
 & \fedra & 1.0 & 1.0 & 0.34 & 1.0 & 1.0 & 0.43 & 1.0 & 1.0 & 0.61\\
  \hline
q9 & FedX & 10.8 & 3.0 & 1.55 & 73.0 & 3.0 & 1.45 & 143.0 & 3.0 & 1.95\\
 & DAW & 12.4 & 2.0 & 0.39 & 27.2 & 1.6 & 0.45 & 47.2 & 1.6 & 0.88\\
 & \fedra & 3.0 & 3.0 & 0.41 & 3.0 & 3.0 & 0.65 & 3.0 & 3.0 & 0.77\\
  \hline
q10 & FedX & 10.8 & 2.0 & 1.48 & 49.2 & 2.0 & 1.33 & 96.4 & 2.0 & 1.81\\
 & DAW & 14.2 & 1.6 & 0.43 & 29.2 & 1.4 & 0.46 & 53.4 & 1.0 & 1.09\\
 & \fedra & 3.0 & 2.0 & 0.53 & 5.2 & 2.0 & 0.84 & 9.2 & 2.0 & 1.29\\
  \hline
q11 & FedX & 21.6 & 2.0 & 1.44 & 43.4 & 2.0 & 1.22 & 94.8 & 2.0 & 1.52\\
 & DAW & 7.2 & 1.0 & 0.36 & 11.6 & 1 & 0.38 & 18.8 & 1.0 & 0.53\\
 & \fedra & 2.0 & 2.0 & 0.35 & 2.0 & 1.0 & 0.43 & 2.0 & 1.0 & 0.57\\
  \hline
q12 & FedX & 10.8 & 4.0 & 1.26 & 98.0 & 4.0 & 1.41 & 192.0 & 4.0 & 1.85\\
 & DAW & 26.4 & 2.6 & 0.63 & 60.4 & 3 & 0.71 & 113.6 & 2.2 & 1.56\\
 & \fedra & 4.0 & 4.0 & 0.59 & 4.0 & 3.0 & 0.76 & 4.0 & 2.0 & 1.62\\
  \hline
q13 & FedX & 32.4 & 3.0 & 1.37 & 73.0 & 3.0 & 1.56 & 143.0 & 3.0 & 1.91\\
 & DAW & 18.0 & 2.0 & 0.53 & 39.6 & 1.6 & 0.51 & 73.0 & 0.8 & 0.87\\
 & \fedra & 3.0 & 3.0 & 0.45 & 3.0 & 3.0 & 0.58 & 3.0 & 2.0 & 1.05\\
  \hline
q14 & FedX & 54.0 & 1.0 & 1.40 & 24.2 & 1.0 & 1.46 & 47.4 & 1.0 & 1.91\\
 & DAW & 4.6 & 0.0 & 0.30 & 9.8 & 0 & 0.31 & 18.6 & 0.0 & 0.40\\
 & \fedra & 1.0 & 1.0 & 0.32 & 1.0 & 1.0 & 0.36 & 1.0 & 0.0 & 0.47\\
  \hline
q15 & FedX & 54.0 & 1.0 & 1.46 & 24.2 & 1.0 & 1.52 & 47.4 & 1.0 & 1.88\\
 & DAW & 5.0 & 0.0 & 0.45 & 12.8 & 0 & 0.75 & 17.0 & 0.0 & 0.93\\
 & \fedra & 1.0 & 1.0 & 0.46 & 1.0 & 1.0 & 0.79 & 1.0 & 0.0 & 1.26\\
  \hline
q16 & FedX & 43.2 & 2.0 & 1.57 & 48.8 & 2.0 & 1.60 & 95.6 & 2.0 & 1.69\\
 & DAW & 13.2 & 2.0 & 0.41 & 28.0 & 1.6 & 0.51 & 52.8 & 1.6 & 0.80\\
 & \fedra & 2.0 & 2.0 & 0.39 & 2.0 & 2.0 & 0.53 & 2.0 & 2.0 & 0.85\\
  \hline
q17 & FedX & 10.8 & 1.0 & 1.58 & 24.2 & 1.0 & 1.53 & 47.4 & 1.0 & 1.72\\
 & DAW & 6.8 & 0.0 & 0.58 & 14.2 & 0 & 0.54 & 24.2 & 0.0 & 1.26\\
 & \fedra & 1.0 & 0.0 & 0.43 & 1.0 & 0.0 & 0.63 & 1.0 & 0.0 & 0.91\\
  \hline
q18 & FedX & 43.2 & 1.0 & 1.44 & 24.2 & 1.0 & 1.49 & 47.4 & 1.0 & 1.69\\
 & DAW & 4.0 & 0.0 & 0.32 & 9.4 & 0 & 0.35 & 15.0 & 0.0 & 0.44\\
 & \fedra & 1.0 & 1.0 & 0.31 & 1.0 & 1.0 & 0.38 & 1.0 & 1.0 & 0.48\\
  \hline
q19 & FedX & 10.8 & 2.0 & 1.19 & 48.8 & 2.0 & 1.34 & 95.6 & 2.0 & 1.53\\
 & DAW & 12.4 & 0.8 & 0.42 & 28.0 & 0.6 & 0.54 & 50.2 & 0.6 & 0.78\\
 & \fedra & 2.0 & 2.0 & 0.39 & 2.0 & 2.0 & 0.51 & 2.0 & 2.0 & 0.92\\
 \hline
 q20 & FedX & 32.4 & 3.0 & 1.34 & 73.0 & 3.0 & 1.00 & 143.0 & 3.0 & 1.85\\
 & DAW & 18.4 & 2.0 & 0.48 & 38.0 & 2 & 0.55 & 69.8 & 1.2 & 1.31\\
 & \fedra & 3.0 & 3.0 & 0.46 & 3.0 & 3.0 & 0.75 & 3.0 & 3.0 & 1.23\\
 \hline
\end{tabular}
}
\label{table:sourceselection}
\end{center}
\end{table}

The benchmark queries were processed using DAW and \fedra source selection strategies
over federations of sizes 10, 25 and 50, and an age limit of one month. 
For each size, five random federations were 
considered, and the average of the results is presented in Table~\ref{table:sourceselection}.
Additionally, 
FedX execution plan was used to compute its number of selected sources. 
We hypothesize that \fedra is able to reduce the number of selected sources and the number of selected public endpoints using
the endpoints descriptions.
%as far as the containment relationship holds among the replicas and the original dataset. 
The results suggest that \fedra is able to reduce the number of selected sources in more than one
 order of magnitude, and the number of selected public endpoints remains low, and the number of
 selected sources remains stable independently of the federation size for \fedra, while it increases
 with the federation size for FedX and DAW. 
 As depicted by the evolution of NSPS for each query, we can see that public endpoint nearly disappear 
 for the two months age limit and consequently, the load on the public endpoint effectively decreases. 
 In general, \fedra selects less sources than DAW, but DAW source selection
 present a greater reduction of the number of times the public endpoint is selected.
 
%The size of the reduction of the number of selected sources is strongly related to the number declared containments among the replicas.
%For replicas that contain little to none containments there is only a slightly reduction, but for richly described replicas where there is an elevated
%containment among the replicas, this reduction can be of up to XX orders of magnitude.

\subsection{Preservation of the Answer}

The selected sources in the previous from the previous part were used to 
annotate SPARQL 1.1 queries using the SERVICE clause, than the
annotated queries were executed using FedX.
We hypothesize that executing the queries produced by \fedra leads to a similar 
number of answers than 
 executing the queries directly by FedX, possibly some stale values may appear since the
used age limit is one month.
 %Moreover, \fedra does not discard any relevant source, then the query execution leads to a
 %complete answer. 
 Table~\ref{table:results} shows the achieved completeness reported by these executions. 
Some query executions present unexpected results, even if the public endpoint was the only one selected (e.g, q12). 
An analysis of the cases were the completeness is
low and the staleness is high evidences that FedX execution present 
some problems when special characters are present in the queries or
when OPTIONALs and FILTERs are used with SERVICE clases.
  %These results suggest that our hypothesis holds, and \fedra is actually able to increase recall. 

\subsection{\fedra Cost}

  \begin{table}[!htbp]
 \begin{center}
\caption{Enhanced scalability thanks to the source selection approaches over different federations sizes}
{
\scriptsize
\begin{tabular}{|c|c|r|r|r|r|r|r|r|r|r|r|r|r|}
 \hline
Query&Approach & \multicolumn{4}{c|}{10 endpoints Federation} & \multicolumn{4}{c|}{25 endpoints Federation} & \multicolumn{4}{c|}{50 endpoints Federation}\\
\cline{3-14}
 & & ETUE & \# Answers & \multicolumn{1}{c|}{C} & \multicolumn{1}{c|}{S} & ETUE & \# Answers & \multicolumn{1}{c|}{C} & \multicolumn{1}{c|}{S} & ETUE & \# Answers & \multicolumn{1}{c|}{C} & \multicolumn{1}{c|}{S}\\
  \hline
q1 & FedX & 1.05 & 3.0 & 1.00 & 0.00 & 1.92 & 3.0 & 1.00 & 0.00 & 2.64 & 3.0 & 1.00 & 0.00\\
 & DAW & 1.58 & 3.0 & 1.00 & 0.00 & 1.23 & 3.0 & 1.00 & 0.00 & 1.70 & 3.0 & 1.00 & 0.00\\
 & \fedra & 0.83 & 3.0 & 1.00 & 0.00 & 1.00 & 3.0 & 1.00 & 0.00 & 1.22 & 3.0 & 1.00 & 0.00\\
  \hline
q2 & FedX & 1.41 & 1.0 & 1.00 & 0.00 & 1.64 & 1.0 & 1.00 & 0.00 & 1.55 & 1.0 & 1.00 & 0.00\\
 & DAW & 1.71 & 1.0 & 1.00 & 0.00 & 1.13 & 1.0 & 1.00 & 0.00 & 0.96 & 1.0 & 1.00 & 0.00\\
 & \fedra & 0.68 & 1.0 & 1.00 & 0.00 & 0.75 & 1.0 & 1.00 & 0.00 & 0.95 & 1.0 & 1.00 & 0.00\\
  \hline
q3 & FedX & 300.00 & 0.0 & 0.00 & 0.00 & 300.00 & 0.0 & 0.00 & 0.00 & 300.00 & 0.0 & 0.00 & 0.00\\
 & DAW & 300.00 & 2381.6 & 0.74 & 0.00 & 300.00 & 6586.6 & 0.80 & 0.00 & 276.95 & 9890.0 & 0.80 & 0.00\\
 & \fedra & 0.60 & 144.0 & 0.20 & 0.00 & 0.74 & 144.0 & 0.20 & 0.00 & 0.92 & 144.0 & 0.20 & 0.00\\
  \hline
q4 & FedX & 3.54 & 421.2 & 1.00 & 0.00 & 14.27 & 943.8 & 1.00 & 0.00 & 48.57 & 1864.2 & 1.00 & 0.00\\
 & DAW & 1.71 & 0.0 & 0.00 & 0.00 & 2.24 & 0.0 & 0.00 & 0.00 & 1.53 & 0.0 & 0.00 & 0.00\\
 & \fedra & 0.75 & 39.0 & 1.00 & 0.00 & 0.76 & 39.0 & 1.00 & 0.00 & 1.19 & 39.0 & 1.00 & 0.00\\
  \hline
q5 & FedX & 300.00 & 22405471.0 & 1.00 & 0.00 & 300.00 & 1662636.8 & 0.20 & 0.00 & 300.00 & 0.0 & 0.00 & 0.00\\
 & DAW & 300.00 & 1675486.8 & 0.60 & 0.00 & 300.00 & 0.0 & 0.00 & 0.00 & 182.54 & 0.0 & 0.00 & 0.00\\
 & \fedra & 0.95 & 20481.4 & 0.60 & 0.00 & 1.08 & 16758.0 & 0.40 & 0.00 & 1.31 & 59584.0 & 0.60 & 0.00\\
  \hline
q6 & FedX & 1.10 & 54.0 & 1.00 & 0.00 & 1.08 & 121.0 & 1.00 & 0.00 & 1.27 & 240.0 & 1.00 & 0.00\\
 & DAW & 1.67 & 50.0 & 1.00 & 0.00 & 2.25 & 115.0 & 1.00 & 0.00 & 2.80 & 216.0 & 1.00 & 0.00\\
 & \fedra & 34.10 & 5.0 & 1.00 & 0.00 & 34.37 & 5.0 & 1.00 & 0.00 & 89.49 & 5.0 & 1.00 & 0.00\\
  \hline
q7 & FedX & 1.32 & 8.0 & 1.00 & 0.00 & 1.11 & 8.0 & 1.00 & 0.00 & 1.31 & 8.0 & 1.00 & 0.00\\
 & DAW & 1.67 & 8.0 & 1.00 & 0.00 & 2.34 & 8.0 & 1.00 & 0.00 & 2.86 & 8.0 & 1.00 & 0.00\\
 & \fedra & 0.57 & 8.0 & 1.00 & 0.00 & 0.83 & 8.0 & 1.00 & 0.00 & 1.11 & 8.0 & 1.00 & 0.00\\
  \hline
q8 & FedX & 1.15 & 54.0 & 1.00 & 0.00 & 0.94 & 121.0 & 1.00 & 0.00 & 1.57 & 240.0 & 1.00 & 0.00\\
 & DAW & 1.59 & 23.0 & 1.00 & 0.00 & 2.59 & 59.0 & 1.00 & 0.00 & 2.32 & 104.0 & 1.00 & 0.00\\
 & \fedra & 0.86 & 5.0 & 1.00 & 0.00 & 1.04 & 5.0 & 1.00 & 0.00 & 1.28 & 5.0 & 1.00 & 0.00\\
  \hline
q9 & FedX & 47.66 & 728524.8 & 1.00 & 1.00 & 300.00 & 4392483.6 & 1.00 & 1.00 & 300.00 & 4731512.4 & 1.00 & 1.00\\
 & DAW & 18.64 & 17625.2 & 1.00 & 1.00 & 156.73 & 48615.0 & 1.00 & 0.80 & 120.45 & 276364.2 & 1.00 & 1.00\\
 & \fedra & 300.00 & 576.0 & 1.00 & 1.00 & 300.00 & 576.0 & 1.00 & 1.00 & 300.00 & 576.0 & 1.00 & 1.00\\
  \hline
q10 & FedX & 300.00 & 1051826.6 & 0.03 & 0.02 & 300.00 & 1015418.2 & 0.02 & 0.01 & 300.00 & 1900207.4 & 0.02 & 0.01\\
 & DAW & 300.00 & 93712.8 & 0.00 & 0.00 & 222.01 & 105893.6 & 0.00 & 0.00 & 165.71 & 52953.6 & 0.00 & 0.00\\
 & \fedra & 0.60 & 17567.0 & 0.00 & 0.00 & 0.81 & 41244.8 & 0.00 & 0.00 & 1.28 & 113765.6 & 0.00 & 0.00\\
  \hline
q11 & FedX & 1.59 & 10.8 & 1.00 & 1.00 & 1.73 & 24.2 & 1.00 & 1.00 & 3.01 & 47.8 & 1.00 & 1.00\\
 & DAW & 1.32 & 63.6 & 1.00 & 1.00 & 1.57 & 138.0 & 1.00 & 1.00 & 3.12 & 350.4 & 1.00 & 1.00\\
 & \fedra & 0.75 & 6.0 & 1.00 & 1.00 & 0.91 & 6.0 & 1.00 & 1.00 & 0.95 & 6.0 & 1.00 & 1.00\\
  \hline
q12 & FedX & 300.00 & 0.0 & 0.00 & 0.00 & 300.00 & 0.0 & 0.00 & 0.00 & 300.00 & 0.0 & 0.00 & 0.00\\
 & DAW & 300.00 & 0.0 & 0.00 & 0.00 & 300.00 & 0.0 & 0.00 & 0.00 & 300.00 & 0.0 & 0.00 & 0.00\\
 & \fedra & 0.71 & 0.0 & 0.00 & 0.00 & 0.68 & 0.0 & 0.00 & 0.00 & 1.23 & 0.0 & 0.00 & 0.00\\
  \hline
q13 & FedX & 16.67 & 2.0 & 1.00 & 0.00 & 214.07 & 2.0 & 1.00 & 0.00 & 300.00 & 2.0 & 1.00 & 0.00\\
 & DAW & 10.51 & 2.0 & 1.00 & 0.00 & 177.39 & 2.0 & 1.00 & 0.00 & 256.30 & 2.0 & 1.00 & 0.00\\
 & \fedra & 261.63 & 2.0 & 1.00 & 0.00 & 278.49 & 2.0 & 1.00 & 0.00 & 300.00 & 2.0 & 1.00 & 0.00\\
  \hline
q14 & FedX & 1.33 & 1.0 & 1.00 & 0.00 & 2.41 & 1.0 & 1.00 & 0.00 & 2.18 & 1.0 & 1.00 & 0.00\\
 & DAW & 1.37 & 1.0 & 1.00 & 0.00 & 1.90 & 1.0 & 1.00 & 0.00 & 2.76 & 1.0 & 1.00 & 0.00\\
 & \fedra & 0.78 & 1.0 & 1.00 & 0.00 & 0.73 & 1.0 & 1.00 & 0.00 & 0.98 & 1.0 & 1.00 & 0.00\\ 
 \hline
q15 & FedX & 1.22 & 43.2 & 1.00 & 0.00 & 2.85 & 96.8 & 1.00 & 0.00 & 1.81 & 191.2 & 1.00 & 0.00\\
 & DAW & 1.64 & 20.0 & 1.00 & 0.00 & 2.01 & 51.2 & 1.00 & 0.00 & 3.29 & 68.0 & 1.00 & 0.00\\
 & \fedra & 32.09 & 4.0 & 1.00 & 0.00 & 32.83 & 4.0 & 1.00 & 0.00 & 36.54 & 4.0 & 1.00 & 0.00\\
  \hline
q16 & FedX & 300.00 & 7142.4 & 0.11 & 0.07 & 300.00 & 2764.4 & 0.04 & 0.06 & 300.00 & 1208.2 & 0.01 & 0.15\\
 & DAW & 300.00 & 5640.0 & 0.08 & 0.06 & 300.00 & 2965.0 & 0.04 & 0.08 & 300.00 & 1402.0 & 0.02 & 0.19\\
 & \fedra & 121.89 & 21224.0 & 0.33 & 0.03 & 181.42 & 17922.0 & 0.28 & 0.03 & 124.27 & 9925.0 & 0.15 & 0.05\\
  \hline
q17 & FedX & 1.37 & 64.8 & 1.00 & 0.00 & 2.72 & 145.2 & 1.00 & 0.00 & 2.83 & 286.8 & 1.00 & 0.00\\
 & DAW & 1.76 & 0.0 & 0.00 & 0.00 & 1.78 & 0.0 & 0.00 & 0.00 & 2.81 & 0.0 & 0.00 & 0.00\\
 & \fedra & 0.67 & 0.0 & 0.00 & 0.00 & 0.68 & 0.0 & 0.00 & 0.00 & 0.94 & 0.0 & 0.00 & 0.00\\
  \hline
q18 & FedX & 1.59 & 345.6 & 1.00 & 0.00 & 300.00 & 774.4 & 1.00 & 0.00 & 300.00 & 1529.6 & 1.00 & 0.00\\
 & DAW & 1.71 & 128.0 & 1.00 & 0.00 & 2.02 & 300.8 & 1.00 & 0.00 & 2.17 & 473.8 & 1.00 & 0.00\\
 & \fedra & 0.63 & 32.0 & 1.00 & 0.00 & 0.88 & 32.0 & 1.00 & 0.00 & 0.94 & 32.0 & 1.00 & 0.00\\
  \hline
q19 & FedX & 2.40 & 19038.4 & 1.00 & 0.00 & 9.71 & 98193.2 & 1.00 & 0.00 & 12.61 & 379105.4 & 1.00 & 0.00\\
 & DAW & 2.88 & 5737.6 & 1.00 & 0.00 & 3.33 & 28977.4 & 1.00 & 0.00 & 8.04 & 53846.6 & 1.00 & 0.00\\
 & \fedra & 0.86 & 163.0 & 1.00 & 0.00 & 0.73 & 163.0 & 1.00 & 0.00 & 1.13 & 163.0 & 1.00 & 0.00\\
  \hline
q20 & FedX & 300.00 & 0.0 & 0.00 & 0.00 & 2.46 & 0.0 & 0.00 & 0.00 & 2.65 & 0.0 & 0.00 & 0.00\\
 & DAW & 9.55 & 12021.8 & 1.00 & 0.00 & 69.25 & 74661.2 & 0.80 & 0.00 & 84.95 & 591487.0 & 1.00 & 0.00\\
 & \fedra & 300.00 & 24.0 & 0.07 & 0.00 & 300.00 & 24.0 & 0.07 & 0.00 & 300.00 & 24.0 & 0.07 & 0.00\\
 \hline
\end{tabular}
}
\label{table:results}
\end{center}
\end{table}

 Tables~\ref{table:sourceselection} and \ref{table:results} present the  source selection time (SST), 
 and execution time (ETUE) for the previous experiment.
 For most of the queries \fedra SST time is slightly greater than DAW but in most of the queries lower than FedX.
 Nevertheless, \fedra reduces the total execution time in more than 75\% of the
 queries and this reduction may be huge. In some queries the execution time actually increases, we compare
 FedX execution plans to understand why, and notice that the plan obtained from a query with SERVICE clauses presents
 less opportunities for FedX optimizations, then maybe another way  of providing the source selection
 to fedX may  be used, like filtering the endpoints file and providing FedX just with the ones that have been selected.

\section{Related Work}
\label{sec:relatedWork}
The Semantic Web community has proposed different approaches to consume Linked Data from federations of 
endpoints~\cite{DBLP:conf/semweb/AcostaVLCR11,DBLP:conf/semweb/BascaB10,DBLP:conf/semweb/GorlitzS11,DBLP:conf/semweb/SchwarteHHSS11}. Although
source selection and query processing techniques have successfully implemented, none of these approaches is able to exploit information about data replication to enhance  performance and answer completeness. 
 Recently Saleem et al. propose DAW~\cite{DBLP:conf/semweb/SaleemNPDH13}, a source selection technique that relies on 
data summarization  to describe RDF replicas and thus,  reduces the number of selected endpoints. For each triple pattern in a SPARQL query, DAW exploits information 
encoded in source summaries to rank relevant sources in terms of how much they can contribute to the answer. 
Source summaries are expressed as min-wise independent permutation vectors (MIPs)  that index all the predicates in 
a source. Although properties of MIPs are exploited to efficiently estimate the overlap of two sources,
since Linked Data can frequently change,  DAW source summaries may need to be 
regularly recomputed to avoid obsolete answers. To overcome this limitation, \fedra provides a more 
abstract description of the sources which is less sensible  to data changes;
% In \fedra, sources are described  
%in terms of views and containment relationships between these views; 
data provenance and timestamps are stored to control divergence. 

 Divergence and data replication, in general,   have been  widely studied in distributed systems and databases.  
 There is a fundamental trade-off between data consistency, availability and tolerance to failures. In one 
 extreme, distributed systems implement the ACID transactional model and ensure strong 
 consistency~\cite{DBLP:journals/computer/Brewer12}. 
 %That is, systems provide atomic, isolated and persistent execution. Changes in the data are transparent to the 
 %users, and distributed systems behave as centralized systems where concurrent updates have the same effects as 
% if they were executed in a serial order. 
Although ACID transactional models have been extensively 
 implemented, it has been shown that in large-scale systems where data is 
 partitioned, ensuring ACID transactions and mutual consistency is not feasible~\cite{DBLP:journals/computer/Brewer12}. 
 %Moreover, the CAP 
 %theorem~\cite{DBLP:journals/computer/Brewer12} states that distributed systems with replicated data that 
 %ensure consistency, availability and partition tolerance at the same time cannot be implemented, i.e., 
 %at most two properties can be perfectly guaranteed. 
 %These results imply that in systems with high-availability, strong consistency may be sacrificed.
 % Finally, in the context of the Semantic Web, the trade-off between  
 %latency and consistency has also gained attention~\cite{DBLP:conf/semweb/Rietveld12,DBLP:conf/esws/Schandl10}, 
 %and approaches to describe changes in the replicas and synchronize local changes once replicas are connected to 
 %the original data source. 
 Based on these results, \fedra does not rely on strong consistency 
 that would constrain Linked Data participants. Contrary, \fedra exploits source  descriptions  to
reduce the number of contacted endpoints while satisfies divergence thresholds. Thus, depending on the 
 divergence tolerated  by a user, a query can be performed against  replicas if  accessing 
the original source is not possible. 

\section{Conclusions and Future Work}
\label{sec:conclusion}
We presented \fedra a source selection approach that takes advantage of replicated data, and reduces the number of selected endpoints given a threshold of divergence. A lower value of divergence allows query engines to produce fresher answers but at the risk of failing to get answers. Contrary, a higher value risks engines to retrieve answers that are
obsolete, while chances of producing answers increases as the space of possible selected sources is larger.  

In the future, we plan to consider a finer-grained granularity of divergence. Instead of considering divergence of a replica, we will compute divergence for specific predicates, \emph{e.g.}, divergence of predicates that represent links between datasets. In addition, we plan to integrate DAW with \fedra to handle replicas with missing descriptions. 

\subsubsection*{Acknowledgments.} We thank Andreas Schwarte, who provided us with a version of FedX that prints the execution plans.

\bibliographystyle{abbrv}
\bibliography{references}

\end{document}